\documentclass{article}
\usepackage{amsmath}
\usepackage{graphicx}
\usepackage{url}
\usepackage[comma,numbers,sort&compress]{natbib}
\usepackage{wrapfig}

\begin{document}

\title{Sudoku as a special transportation problem}
\author{%
Mansour Moufid%
\footnote{\texttt{mansour@peripetylabs.com}}%
}
\date{July 2010}
\maketitle

\begin{abstract}
Sudoku is a popular combinatorial puzzle.
A new method of solving Sudoku is presented, which involves formulating a
puzzle as a special type of transportation problem.
This model allows one to solve puzzles with more than one solution,
keeping the constraints of the problem fixed,
and simply changing a cost matrix between solutions.
\end{abstract}

\section{Introduction}

Sudoku is a popular combinatorial puzzle consisting of a $9 \times 9$,
partially-filled grid of digits 1 to 9. \cite{unwed}
The remainder of the grid must be filled in according to the rules:

\begin{itemize}
  \item[I.] every row contains only one of each digit 1--9;
  \item[II.] every column contains only one of each digit; and
  \item[III.] every one of the nine $3 \times 3$ sub-grids contains
    only one of each digit.
\end{itemize}

Sudoku has been modeled and solved using a wide variety of methods,
for example, as an exact cover problem solved using the famous
Dancing Links algorithm. \cite{knuth}
Examples from the field of operations research include the binary integer
linear program,%
\footnote{%
Not to be confused with ``binary Sudoku'': \url{http://xkcd.com/74/}
}
which involves 729 ($81 \times 9$) binary variables representing the state
of each of the 81 places in the square. \cite{bartlett}
This is equivalent to a constraint satisfaction problem,
where any feasible solution is a solution. \cite{simonis}

The latter methods are equivalent to the general transportation problem.
However, for problems with more than one solution,
these methods require iteratively reformulating the problem by inserting
additional constraints until no new solution is found.
The method proposed here differs from these in that the transportation
problem provides an intuitive and compact representation for solving problems
with many solutions.
This method distinguishes between different solutions to a given problem
by manipulating the cost matrix,
without otherwise modifying the constraints of the problem.

\section{Sudoku as a transportation problem}

\subsection{A general transportation model}

Transportation problems are a special type of linear programming problem.
\cite[Ch.~14, p.~299]{transport}
A transportation problem generally consists of:

\begin{itemize}
  \item a set of $m$ supply points,
        $\boldsymbol{s} = \left\{s_i;\,i=1,\ldots,m\right\}$;
  \item a set of $n$ demand points,
        $\boldsymbol{d} = \left\{d_j;\,j=1,\ldots,n\right\}$; and
  \item a variable cost matrix,
        $\boldsymbol{c} =
         \left\{c_{i,j};\,i=1,\ldots,m,\,j=1,\ldots,n\right\}$,
        representing the cost of transporting one unit
        from each supply point to every demand point.
\end{itemize}

The objective is to minimize the total cost of shipments while meeting demand.
Let $x_{i,j}$ be the number of units shipped from a supply point $i$
to a demand point $j$.
Then a transportation problem has the general formulation:
\cite[\S7.1]{transportlp}

\begin{align}
\label{lp}
  \text{minimize}\quad
    z &= \sum_{i = 1}^m \sum_{j = 1}^n c_{i,j} x_{i,j} \\
  \text{subject to}\quad
    \sum_{j = 1}^n x_{i,j} &\le s_i
                           & \forall i = 1, \ldots, m \nonumber \\
    \sum_{i = 1}^m x_{i,j} &\ge d_j
                           & \forall j = 1, \ldots, n \nonumber \\
    x_{i,j}                &\ge 0
                           & \forall i = 1, \ldots, m \text{;}\,
                                     j = 1, \ldots, n \nonumber
\end{align}

If the total supply of goods is equal to the total demand,
that is $\sum_i s_i = \sum_j d_j$,
then the problem is said to be balanced.

\begin{figure}
\label{fig:network}
\begin{center}
  \includegraphics{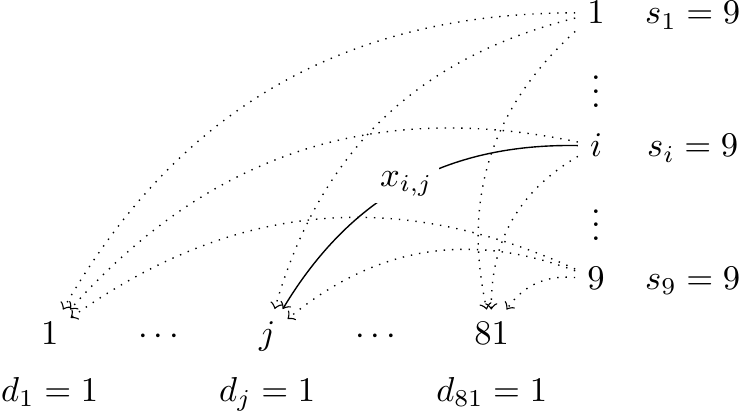}
\end{center}
\caption{%
A transportation network model of a Sudoku puzzle.
There are nine supply points, with a supply of nine units each; and eighty one
demand points, with a demand of one unit each.
Each one of the nine supply points, $i$, can ship to every one of the eighty
one demand points, $j$, one unit at a cost of $c_{i,j}$.
}
\end{figure}

Solving a Sudoku puzzle involves placing (or transporting) up to
eighty one digits into an equal number of cells.
Therefore, in modeling a Sudoku puzzle as a transportation problem, there are
eighty one demand points to consider, each with a demand of one unit.
The eighty one digits available to be placed in the grid consist of (nine of
each of) the digits one through nine.
Thus, there are nine units in each of nine distinct supply points to be
considered.
That is, $m = 9$ and $n = 81$ in Equations~\ref{lp}.
Let $\boldsymbol{s}$ be the set of supply points,
and let $\boldsymbol{d}$ be the set of demand points, such that
$s_i = 9$ for all $i = 1, \ldots, 9$ and $d_j = 1$ for all
$j = 1, \ldots, 81$.
An illustration of this transportation network model of Sudoku is shown
in Figure~\ref{fig:network}.

\subsection{Transportation cost matrix}

In modeling different Sudoku puzzles, the supply and demand sets
$\boldsymbol{s}$ and $\boldsymbol{d}$ remain constant.
A variable cost, $c_{i,j}$, is incurred when transporting one unit from a
supply point $i$ to a demand point $j$.
Thus, it is necessary to formulate a cost matrix $c$ which assigns a cost of
transporting one unit from supply point to demand point.

Let a puzzle matrix, $\Pi$, be a $9 \times 9$ matrix representing the
contents of the initial Sudoku puzzle grid.
An element of the matrix $\Pi$ is non-zero if its corresponding cell in the
Sudoku grid contains a digit, in which case its value is equal to that of the
digit in the cell, and zero otherwise.
Let $\pi_{i, j}$ represent the elements of the puzzle matrix $\Pi$, where
$i = 1, \ldots, 9$ and $j = 1, \ldots, 9$.

\begin{equation*}
\Pi = \left[
  \begin{array}{ccc}
    \pi_{1, 1} & \cdots & \pi_{1, 9} \\
    \vdots     & \ddots & \vdots     \\
    \pi_{9, 1} & \cdots & \pi_{9, 9}
  \end{array}
\right]
\end{equation*}

Each element in this matrix has an integer value $\lambda = 0, 1, \ldots, 9$.
Given an initial puzzle matrix $\Pi$, we seek $c_{i,j}$ for all
$i = 1, \ldots, 9$ and $j = 1, \ldots, 81$.
Consider one element in a given puzzle matrix, $\pi_{a, b} = \lambda$.
If $\lambda$ is non-zero, then according to rules I and II,
the following elements of $\Pi$ cannot then also have a value of $\lambda$:
all elements $\left( a, j \right)$, $j = 1, \ldots, 9 \neq b$;
and all elements $\left( i, b \right)$, $i = 1, \ldots, 9 \neq a$.
Furthermore, consider that for any coordinate pair $\left( a, b \right)$ where
$a = 1, \ldots, 9$ and $b = 1, \ldots, 9$, there exist $p$, $q$, $r$, and $s$,
such that $a = 3 p + q$, where $p = \left\lfloor a/3 \right\rfloor$,
$q = a \mod 3$; and $b = 3 r + s$, where
$r = \left\lfloor b/3 \right\rfloor$, $s = b \mod 3$.
Thus, according to rule III, $\lambda$ cannot also be the value of
any element $\left( 3 p + m, 3 r + n \right)$ for all $m = 1, 2, 3$
and $n = 1, 2, 3$, where $p = \left\lfloor a/3 \right\rfloor$,
$r = \left\lfloor b/3 \right\rfloor$, and where $3 p + m \neq a$ and
$3 r + n \neq b$.

Thus, given some non-zero element in $\Pi$, $\pi_{a, b} = \lambda \neq 0$,
let $R\left(a, b\right)$ be defined as the set of all
other elements in $\Pi$ whose value cannot also be equal to $\lambda$,
as determined by rules I to III:

\begin{equation}
  R\left(a, b\right) =
  \left\{
    \left( i, b \right) ,\,
    \left( a, j \right) ,\,
    \left(
      3 \left\lfloor \frac{a}{3} \right\rfloor + m,\,
      3 \left\lfloor \frac{b}{3} \right\rfloor + n
    \right)
  \right\}
  \setminus \left\{ \left( a, b \right) \right\}
\end{equation}

where $i = 1, \ldots, 9$, $j = 1, \ldots, 9$,
$m = 1, 2, 3$, and $n = 1, 2, 3$.
This set represents the set of all cells in the same row and column,
and all cells within the same $3 \times 3$ sub-matrix,
as $\pi_{a, b}$, excluding itself.
Let this set be called the \emph{restricted set}.

An illustration of the restricted set for a single digit in a Sudoku grid
is shown below;
in this example, $\pi_{4,4} = 3$ and the corresponding $R\left(4,4\right)$
is shown on the right.

\begin{wrapfigure}[]{r}{2.5in}
\begin{center}
\includegraphics[width=1in]{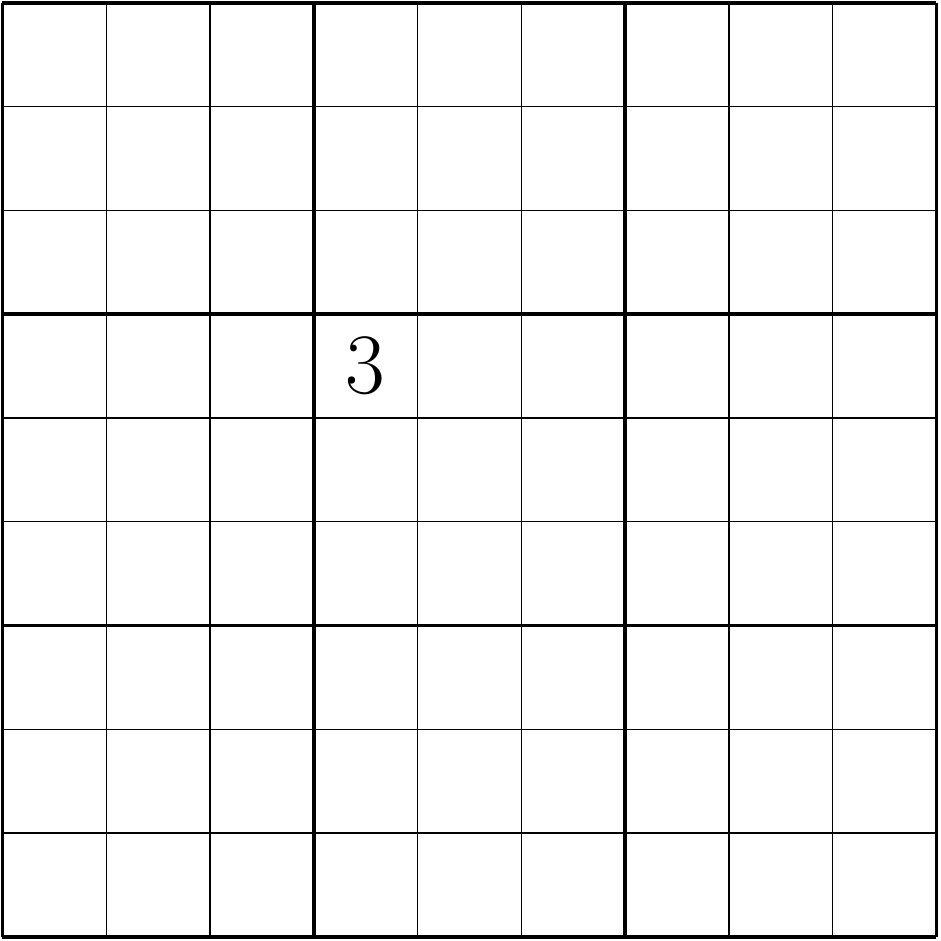}
\includegraphics[width=1in]{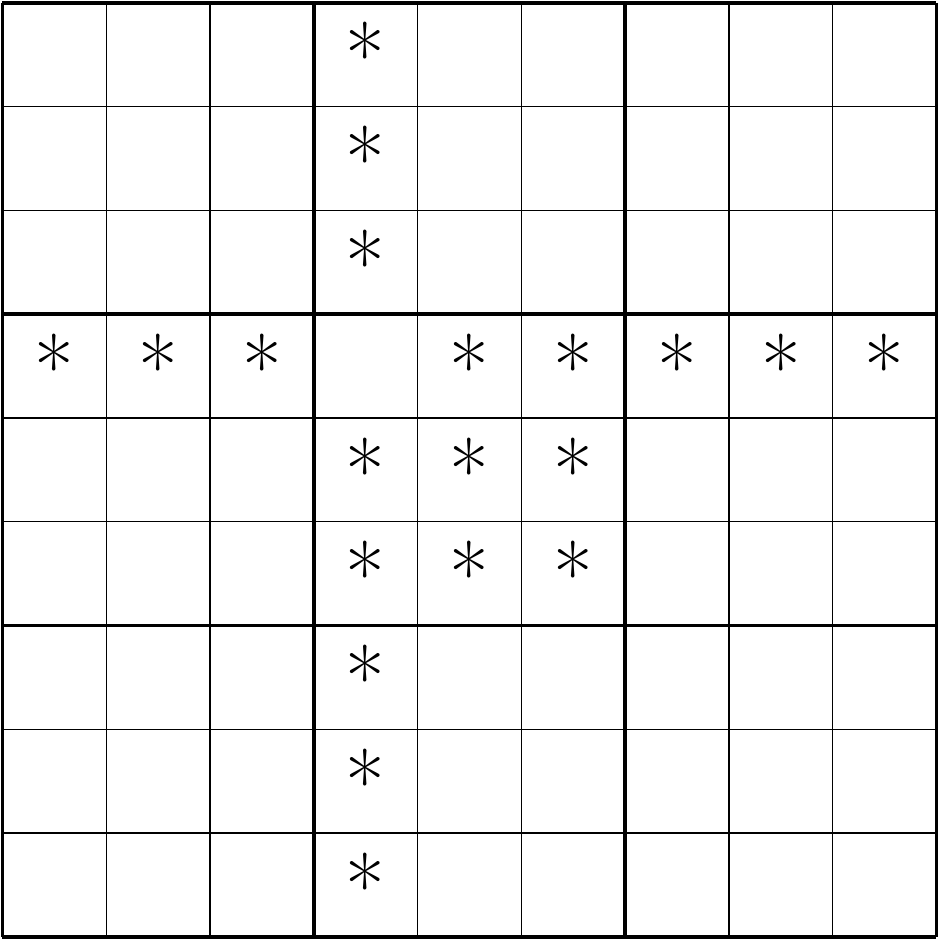}
\end{center}
\end{wrapfigure}

There exist as many restricted sets as there are non-zero elements in $\Pi$.
The intersection of all the restricted sets for any given $\Pi$ can be
expressed in the form of a matrix,
which will be used to formulate the cost matrix.

Consider all non-zero elements of the puzzle matrix $\Pi$,
$\pi_{a,b} = \lambda \neq 0$.
For each of these, let $\phi^{\left( a, b \right)}$ be a matrix of dimensions
$9 \times 9$, defined as:

\begin{equation}
\label{phi1}
  \phi^{\left( a, b \right)}_{i, j} =
  \begin{cases}
    -M & \text{if } \left(i,j\right) = \left(a,b\right)
      \text{;} \\
    M & \text{if } \left( i, j \right) \in R\left(a, b\right)
      \text{; and} \\
    0 & \text{otherwise;}
  \end{cases}
\end{equation}

where $i = 1, \ldots, 9$, $j = 1, \ldots, 9$, and $M > 0$.

Consider the nine possible values, $\lambda = 1,\ldots,9$, of each non-zero
element of $\Pi$.
For each, let $\phi^{\left( \lambda \right)}$ be a matrix of dimensions
$9 \times 9$, defined as:

\begin{equation}
\label{phi2}
  \phi^{\left( \lambda \right)} =
  \sum_{\left( a, b \right) : \pi_{a, b} = \lambda}
  \phi^{\left( a, b \right)}
  \text{.}
\end{equation}

Finally, let $\boldsymbol{r}_i$ represent the row vectors of
${\phi^{\left( \lambda \right)}}$, where $i = 1, \ldots, 9$, and let
$\boldsymbol{\Phi}^{\left( \lambda \right)}$ be the vector defined as:

\begin{equation}
\label{phi4}
  \boldsymbol{\Phi}^{\left( \lambda \right)} = \left[
  \begin{array}{ccc}
    \boldsymbol{r}_1 & \cdots & \boldsymbol{r}_9
  \end{array}
  \right]
  \text{.}
\end{equation}

The variable cost matrix sought, $c$, is given by:

\begin{equation}
\label{costmatrix}
  c = \left[
  \begin{array}{c}
    \boldsymbol{\Phi}^{\left( 1 \right)} \\
    \vdots \\
    \boldsymbol{\Phi}^{\left( 9 \right)} \\
  \end{array}
  \right]
  \text{.}
\end{equation}

Consider again Equation~\ref{phi1}.
A cost of $M$ means that it is highly undesirable to ship a unit to a
particular demand point.
Conversely, a cost of $-M$ means that the shipment is highly desirable.
A cost of zero means that the shipment is possible but without preference over
other possible destinations that also incur a zero cost.
To understand the meaning of $M$ intuitively, consider the limit
$M \rightarrow \infty$.
If $c_{i,j} = M$, this means supply point $i$ can not ship to demand point $j$
``at any cost.''
Conversely, if $c_{i,j} = -M$, this means supply point $i$ must ship to demand
point $j$ ``at all cost.''
In practice, the value of $M$ need only be strictly positive.
Algorithms that solve transportation problems are guaranteed
to find an optimal solution if one exists.
An optimal solution will not assign a shipment from a supply to a demand point
when the cost of such a shipment is $M$; such shipments will all be avoided.
In fact, any optimal solution will only contain shipments between points that
incur zero cost.
For simplicity, let $M = 1$.

There can be more than one optimal solution to the general transportation
problem posed by the supply vector $s$,
the demand vector $d$,
and the cost matrix $c$.
These solutions all have an objective function value of $z = - N M$, where $N$
is the number of digits initially given in the Sudoku puzzle.
However, these solutions do not all form Latin squares, much less valid
solutions to the Sudoku puzzle.
If the Sudoku puzzle has a unique solution, then exactly one of the optimal
solutions to the general transportation problem will also be the solution to
the puzzle.
Therefore, additional constraints to those in the general transportation
problem are needed.

\subsection{A special transportation problem}

We formulate rules I to III as additional constraints,
to obtain the special transportation problem:

\begin{align}
\label{speciallp}
  \text{minimize}\quad
    z &= \sum_{i = 1}^m \sum_{j = 1}^n c_{i,j} x_{i,j} \\
  \text{subject to}\quad
    \sum_{j = 1}^n x_{i,j} &= s_i & i = 1, \ldots, m \nonumber \\
    \sum_{i = 1}^m x_{i,j} &= d_j & j = 1, \ldots, n \nonumber \\
    \sum_{j = 1}^n x_{i,j} &\le 9   & i = 1, \ldots, m \nonumber \\
    \sum_{i = 1}^m x_{i,j} &\le 1   & j = 1, \ldots, n \nonumber \\
    x_{i,j}                &\in \left\{0,1\right\}
    & i = 1, \ldots, m \text{;}\, j = 1, \ldots, n \nonumber
\end{align}

It is now possible to solve the Sudoku puzzle by solving the transportation
problem posed by the supply vector $s$, the demand vector $d$, and the cost
matrix $c$ as defined above, and these additional constraints, using any one
of well-known algorithms, such as the simplex algorithm.
This is an efficient way of solving Sudoku puzzles.

\section{Conclusions}

It has been shown that a Sudoku puzzle can be modeled as a special type of
transportation problem.
The model allows to efficiently solve for all solutions to any given problem,
by keeping constraints fixed and manipulating the cost matrix.
Algorithms for solving transportation problems are well established and
have the attractive property of generally completing in polynomial time.

Sudoku is also a special case of a greater class of problems --
that of completing partially-filled Latin squares --
which has applications in cryptography.

\bibliographystyle{unsrtnat}
\bibliography{sudoku}

\end{document}